	\documentclass[apj,numberedappendix]{emulateapj}

\usepackage{graphics,epsf}
\usepackage{amsmath}                
\usepackage{amsfonts}               
\usepackage{amssymb}                
\usepackage{epsfig}                 
\usepackage{float}
\usepackage{color}
\usepackage{tabularx}

\def \cm{~\rm{cm}}
\def \s{~\rm{s}}
\def \km{~\rm{km}}

\def \K{~\rm{K}}

\def \g{~\rm{g}}
\def \G{~\rm{G}}

\def \erg{~\rm{erg}}

\usepackage{xcolor}
\definecolor{redak}{rgb}{0.9,0.15,0.05}

\begin{document}


\title{The common envelope jets supernova (CEJSN)  r-process scenario}

\author{Aldana Grichener\altaffilmark{1}, Noam Soker\altaffilmark{1,2} }

\altaffiltext{1}{Department of Physics, Technion -- Israel Institute of Technology, Haifa
32000, Israel; aldanag@campus.technion.ac.il; soker@physics.technion.ac.il}
\altaffiltext{2}{Guangdong Technion Israel Institute of Technology, Shantou 515069, Guangdong Province, China}

\begin{abstract}
We study r-process feasibility inside jets launched by a cold neutron star (NS) spiralling-in inside the core of a giant star, and find that such common envelope jets supernova events might be a significant source of heavy r-process elements in the early Universe. We run the stellar evolution code MESA to follow the evolution of low metalicity giant stars that swallow NSs during their late expansion phases and find that in some of the cases the NSs penetrate the core. The Bondi-Hoyle-Lyttleton (BHL) mass accretion rate onto a NS as it spirals-in inside the core is sufficiently high to obtain a neutron rich ejecta as required for the heavy r-process where the second and third r-process elements are synthesized. Due to the small radius of the NS the accretion is through an accretion disk and the outflow is in jets (or bipolar disk winds). The r-process nucleosynthesis takes place inside the jets. 
To account for the r-process abundances in the Galaxy we require that one in ten cases of a NS entering the envelope of a giant star ends as a CEJSN r-process event. 
\end{abstract}

\section{INTRODUCTION}
\label{sec:intro}

One of the topics of intensive research in astrophysics concerns the sites where r-process nucleosynthesis takes place, and in particular the sites of ``heavy r-process'' nucleosynthesis where elements of atomic weight $A \ga 130$, are formed.
Over the years, two of the main contenders have been jets launched by a newly born rapidly rotating neutron star (NS) during a core collapse supernova (CCSN) explosion, termed MHD-driven supernovae (e.g.,   \citealt{Winteleretal2012, HaleviMosta2018}; see \citealt{Thielemannetal2018}  for a review), and the merger of two NSs (e.g., \citealt{Qian2012, Rosswogetal2013}; see \citealt{Thielemannetal2017} for a review). Other scenarios exist as well. \citet{Fryer2006}, for instance, consider r-process in a wind blown by the newly born NS in CCSNe (for a paper on proton rich nucleosynthesis in neutrino-driven wind see, \citealt{Blissetal2018}). In a recent paper \cite{Siegeletal2019} suggest that a collapsar, a black hole formed from a collapsing rotating massive star, launches neutron-rich outflows that lead to r-process nucleosynthesis. 

\cite{Papishetal2015} proposed that jets launched by a NS during a common envelope jets supernova (CEJSN) explosion can also serve as a site of heavy r-process nucleosynthesis. 
In the CEJSNe event a NS companion that spirals-in inside the envelope of a giant star and then through its core explodes the giant by launching jets (e.g., \citealt{SokerGilkis2018} for a recent paper).
The subject of the present study is the CEJSN as a possible nucleosynthesis site for r-process elements in a scenario that we term \textit{The CEJSN r-process scenario}. We describe the basic scenario in section \ref{sec:scenario}.
We note that if a NS energizes a bright event by launching jets in the envelope of a giant star but it does not penetrate its core, then the event is termed a CEJSN impostor \citep{Gilkisetal2019}.

Let us elaborate on the main possible sites for r-process nucleosynthesis.
In the MHD-driven supernova scenario magnetic fields lift material from very close to the NS allowing the neutron fraction to be large \citep{Winteleretal2012}. The high neutrino luminosity from the newly born NS ($L_\nu \approx 5\times 10^{52} \erg \s^{-1}$), on the other hand, lowers the neutron fraction by the interaction $n+\nu_e \rightarrow p + e^{-}$ (e.g., \citealt{Pruet2006, Fischer2010}). \cite{Mostaetal2018} find in their 3D magnetorotationally simulations that for the CCSNe to be a robust site for heavy r-process nucleosynthesis, the magnetic field in the pre-collapse source should be unrealistically large, $\approx 10^{13} \G$. 
In the CEJSN r-process scenario, in contrast, the NS is old and cold and hence the low neutrino luminosity does not turn many neutrons back to protons inside the jets. 

A key property of heavy r-process nucleosynthesis is that it must occur on scarce events. 
Light r-process elements exist in all low-metallicity stars, implying that r-process nucleosynthesis in which these elements are formed took place  continuously from the early times of the Galaxy (e.g., \citealt{Sneden2008}). 
However, the abundance of heavy r-process elements in low-metallicity stars shows very large variations among different stars (e.g., \citealt{Tsujimotoetal2017, Hansenetal2018}), implying that they are synthesized in rare occasions (e.g., \citealt{Qian2000, Argast2004}).

One such rare event is the merger of two NSs (e.g., \citealt{Beniaminietal2016a, Metzger2017, Smarttetal2017, Beniaminietal2018, Dugganetal2018, Holmbecketal2018, Radiceetal2019}), also termed kilonova. The kilonova ejects neutron-rich material in several components, including dynamic ejecta, winds from a (metastable) remnant neutron star, and winds from a post-merger accretion disk. These components differ in their dynamical properties and their nucleosynthesis products. The viewing angle and relative intensities of these outflows determine the observed properties of the kilonova. Although r-process nucleosynthesis takes place in all components,  neutrino reduce the neutron-fraction in the polar outflows preventing the formation of the heaviest elements there. This means that the heavy r-process nucleosynthesis tends to occur in the equatorial outflow (e.g., \citealt{Metzger2017} and references therein).

In a new study \cite{Holmbecketal2018} conclude that actinides are over produced in the dynamical ejecta of binary NS mergers. They argue that there must be a significant contribution of lanthanide-rich and actinide-poor nucleosynthesis from another component, like a disk wind, in the merger event.

However, there are two issues concerning the kilonova as a heavy r-process site. 
The first is that it is not clear whether the kilonova can explain the heavy r-process nucleosynthesis at the very early Galactic evolution (e.g., reviews by \citealt{Thielemannetal2017} and \citealt{Hotokezakaetal2018} and references therein).A binary NS merger is preceded by two CCSNe followed by a time delay due to the gravitational wave inspiral phase. \cite{Bonettietal2018} claim that it is possible that merger events occur early enough in the Galactic history through fast mergers of compact binaries in triple systems. Further research is required to make a conclusive determination whether those triple systems can easily form and if they are common enough to account for the r-process abundances and distributions found in observations.  
A second problem arises from the iron abundances observed in low metalicity stars. The two CCSNe produce non-negligible amounts of iron, so that it is not clear that the NSs merger scenario can account for heavy r-process elements in very iron poor stars. One possible way out of this problem is to consider the cases where in the second explosion the NS has a large natal kick but the binary NS system survives, as might happen in rare occasions \citep{BeniaminiPiran2016, Hotokezakaetal2018}. The merger then takes place outside the iron-enriched medium. Moreover, \cite{Safarzadehetal2019} found that double NSs with rather large natal kicks but very short merging timescales can contribute to r-process enrichment in ultra faint galaxies. Nevertheless, it is uncertain whether NSs mergers that are preceded by a natal kick are common enough to explain the abundances of heavy r-process elements in the early Universe. 
We note on that count that the CEJSN is preceded by only one CCSN, and there is no phase of gravitational wave inspiral.

Another unanswered question is whether the binary NS merger event GW170817 really formed heavy r-process elements. \cite{Waxmanetal2017} argue that if the opacity in the observed binary NS merger GW170817 is provided entirely by Lanthanides, then their deduced mass fraction is about 30 times lower than the fraction of Lanthanides in the solar abundance. 
\cite{LiLiuYuZhang2018} also find a low opacity in GW170817 that cannot be compatible with the suggestion that radioactive r-process elements powered the optical emission in that event. Indeed, it is possible that the kilonova GW170817 was mainly powered by fall back accretion onto the central object rather than by radioactive elements (e.g, \citealt{Matsumotoetal2018}). 

The existence of a single r-process site that accounts for the production of all the heavy elements is a debatable matter. In a recent paper, \cite{JiFrebel2018} argue that a single r-process site produces both the rare earth elements (Lanthanides) and the third r-process peak (heavy r-process).  \cite{Tsujimotoetal2017}, on the other hand, believe it might be possible that more than one r-process site accounts for the formation of heavy r-process elements.

Motivated by the unsettled issues with the kilonova (e.g., \citealt{Coteetal2019}) and the claims several r-process sites, we aim to study in more detail the CEJSN r-process scenario as proposed by \cite{Papishetal2015}.

\section{THE CEJSN R-PROCESS SCENARIO}
\label{sec:scenario}

\begin{figure*}
\begin{center}
\includegraphics[width=1\textwidth]{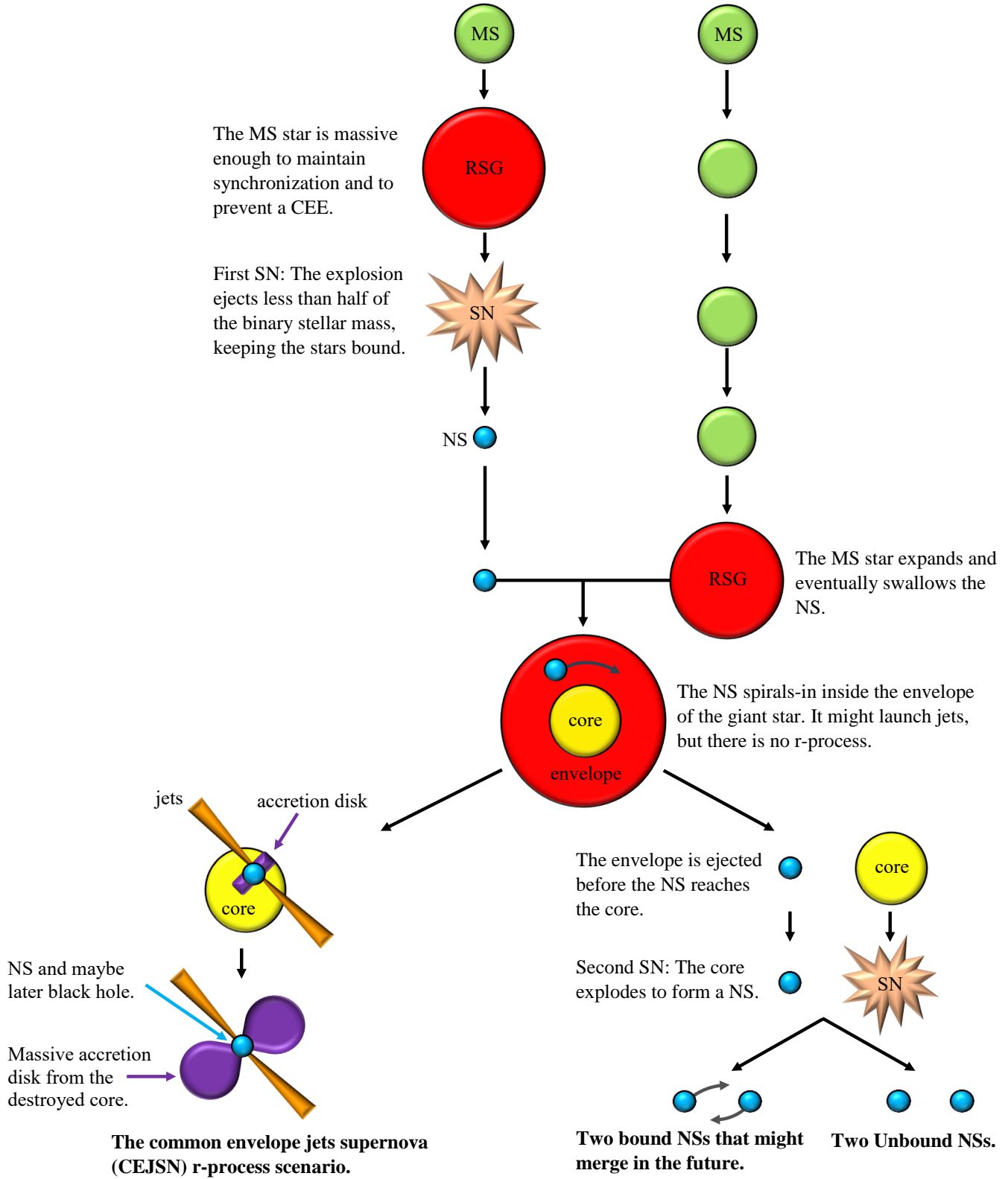}
\vspace*{-2.2cm}
\caption{Possible evolutionary routes of a massive binary stellar system. The initially less massive star swallows the NS during its late expansion phase. The paths diverge according to the location of the NS at the end of the common envelope evolution (CEE). The bottom left describes the CEJSN r-process scenario in which the NS falls into the core. The bottom right describes the evolution towards a double NS binary system. In cases where the NSs remain bound, they might merge at a later stage. }
\label{fig:CEJSN}
\end{center}
\end{figure*}

The evolution of the CEJSN r-process scenario proceeds as follows \citep{Papishetal2015}.
It begins with a detached binary system of two massive main sequence (MS) stars, as we present in Fig. \ref{fig:CEJSN}. The more massive star, plotted in the upper left of Fig. \ref{fig:CEJSN}, evolves to a red super-giant (RSG) and  explodes to form a NS. It might transfer mass to the other star during its RSG phase. Once the initially less massive star expands to become a RSG it can swallow the NS and the system might evolve toward a CEJSN. 
In the CEJSN r-process scenario we assume that the NS penetrates the envelope of the giant star when the giant expands after the exhaustion of helium in its core. At this stage the core is denser than during the early expansion phase, and as we show in section \ref{sec:AcrretionRate}, the accretion rate onto the NS is sufficiently high for heavy r-process nucleosynthesis to occur. 

After the NS enters the envelope of the giant star it begins to spiral-in and to accrete mass from the envelope. The NS might launch jets while it is inside the envelope. At this stage the accretion rate is too low for any r-process nucleosynthesis to take place. Eventually, the spiraling-in process releases enough energy to strip the NS-core system from all or most of the envelope, leading to one of three main outcomes.   

In cases where the envelope is ejected before the NS reaches the core of the giant star the core later explodes as a CCSN to form another NS, as presented in the bottom right panel of Fig. \ref{fig:CEJSN}. Both NSs might become unbound, or else they might form a double NS binary system that can lead to a future merger by emission of gravitational waves (e.g., \citealt{VignaGomezeal2018}). 

The proceeding of the evolution towards the CEJSN r-process scenario is presented in the two bottom left panels of figure \ref{fig:CEJSN}. In this case the NS enters to the remaining core of the giant star.  Note that before the NS enters the core it manages to eject the entire envelope, or at least clean the polar directions, so that the jets are now free to expand out. The NS starts accreting mass from the core through an accretion disk which is likely to launch jets. Eventually the NS destroys the core and part of the core material forms a thick accretion disk around the NS while the rest of the core mass is ejected. Most of the disk mass is accreted onto the NS, that might become then a black hole, and about $10\%$ of the mass leaves the systems in jets (or disk wind). 
  
The destroyed core forms a thick accretion disk with a mass of $M_{\rm disk} \approx 1 M_\odot$ that launches two opposite jets (or bipolar disk wind) with a total mass of $M_{\rm 2j} \approx 0.1 M_\odot$. The fast jets catch-up with the previously ejected core material and interact with it at a distance of about $r_{\rm j,int} \approx 1 R_\odot$ from the NS. This phase lasts for about tens to hundreds of seconds, with a short period of very high mass accretion rate, during which r-process nucleosynthesis takes place (section \ref{sec:AcrretionRate}).

Although some studies argue that it is hard to form an accretion disk inside an envelope (e.g., \citealt{MacLeod2017}), other show that even around main sequence stars, that are much larger than NSs, accretion disks can be formed during the common envelope evolution (e.g., \citealt{Chamandyetal2018}). \cite{Gilkisetal2019} discuss the formation of accretion disks in CEJSNe in detail.

Since in our scenario the core is destroyed when the orbital separation is ${\rm few} \times 0.01 R_\odot$ and its mass is about equal to the mass of the NS, the specific angular momentum of the core-NS binary system at this stage is $\approx 10^{17} \cm^2 \s^{-1}$. This implies that the destructed material of the core has enough angular momentum to form an accretion disk extending out to about a hundred times the NS radius. However, only full 3D hydrodynamical simulations can reveal the true nature of he final interaction and accretion process.

In principle, r-process nucleosynthesis can occur inside jets launched by an NS \citep{Cameron2001, Winteleretal2012, Nishimuraetal2017} or in the post shock jets' material  \citep{PapishSoker2012}. \cite{Papishetal2015} find the post-shock temperature in CEJSNe to be too low for the r-process nucleosynthesis, and conclude that the most promising site for r-process in the CEJSN r-process scenario is inside the jets as they are launched from the NS vicinity. The last phase, in which the already destroyed core forms a disk around the NS, is the phase in which most of the r-process takes place. 
In section \ref{sec:AcrretionRate} we will use the undisturbed core to calculate the accretion rate keeping in mind that this is a crude approximation. 

Most of the ingredients of the CEJSN r-process scenario were studied as separated processes, e.g., the spiraling-in of a NS that launches jets inside a giant envelope (\citealt{Armitage2000, Chevalier2012}), an accretion disk around a NS or a black hole that launches neutron-rich jets (e.g., \citealt{Surman2004, kohri2005}), or more generally the launching of jets by an NS that accretes mass at a high rate \citep{Fryer1996}. Moreover, earlier researches study r-process nucleosynthesis inside jets or inside the hot bubbles they inflate (e.g., \citealt{Fryer2006, Cameron2001, PapishSoker2012}). 

\cite{Papishetal2015} proposed and constructed the CEJSN r-process scenario by considering the operation of these different processes in a coherent single evolution of a binary system composed of an NS and a giant star. They estimated that the mass in the jets is $0.01-0.1 M_\odot$ and the mass synthesized in r-process inside the jets is $\approx 0.001-0.01 M_\odot$. This ratio of r-process to total mass in the jets comes from the results of \cite{Nishimura2006} who found that about $10\%$ of the mass in the jets of CCSNe ends as r-process elements. \cite{Chevalier2012} estimated the rate of events in which the NS enters the envelope of a giant from the results of \cite{Podsiadlowski1995} to be about $1 \%$ of the rate of CCSNe. This leads to the conclusion that not only the jets in the CEJSN r-process scenario have the right conditions for heavy r-process nucleosynthesis, but also the rate of CEJSNe can be compatible with the rate expected in order to explain the r-process abundances, as will be further explained in section \ref{sec:AcrretionRate}.

\cite{CaseySchlaufman2017} claim that the origin of the r-process elements in a very low metalicity star that they study is a Population III or extreme Population II core-collapse supernova that exploded shortly after star formation. Here as well we note that CEJSNe can take place in Population III stars with a very short time delay to explosion. 

In the present study we further explore the CEJSN r-process scenario. By following the evolution of two stellar models (section \ref{sec:NumericalScheme}) we examine the likelihood of the NS to spiral-in into the core (section \ref{sec:FinalOrb}) and we estimate the mass accretion rate onto the NS (section \ref{sec:AcrretionRate}). Our findings, that we summarize in section \ref{sec:summary}, show that the CEJSN is indeed a potentially important r-process site. 

\section{THE EVOLUTIONARY SCHEME}
\label{sec:NumericalScheme}

We use the stellar evolution code MESA (Modules for Experiments in Stellar Astrophysics, e.g \citealt{Paxtonetal2018}) version 9575 to follow the evolution of massive stars in the early universe. We run two non-rotating stellar models with a metalicity of $Z=0.0001$ and initial masses on the zero age main sequence (ZAMS) of $M_{\rm ZAMS}=15M_{\rm \odot}$ and $M_{\rm ZAMS}=30M_{\rm \odot}$. Massive stars suffer from two expansion phases. The first swelling takes place after the exhaustion of hydrogen in the core. A later and more significant swelling occurs when the helium in the core is depleted. At this point the core is consisted of carbon and oxygen (CO). We note that in some cases, depending on metalicity,  mass, and rotation, the star expands monotonically between these two phases (e.g., \citealt{Georgyetal2013}). In our models the stars experience a small contraction between them. We refer to the early expansion phase as the first expansion and to the later expansion phase as the second expansion.    

We assume that the giant star swallows the NS companion of mass $M_{\rm NS}=1.4M_{\rm \odot}$ the second time it expands, initiating a common envelope evolution (CEE) phase in which the NS spirals-in inside the envelope of the giant. Tidal forces will bring the NS into the envelope even if the orbital separation is up to about three times the maximum radius of the giant star (e.g., scaling the expression given by \citealt{Soker1996}).  We mimic the effect of the spiraling-in NS on the envelope of the giant star by removing mass from the envelope of the giant. We examine three possible evolutionary times for the onset of the CEE. 

Let us elaborate first on our $M_{\rm ZAMS}=15M_{\rm \odot}$ stellar model. When the star leaves the main sequence its envelope starts to expand. After the first expansion the radius of the star reaches a value of $R=43R_{\rm \odot}$, and its central density is $\rho \approx 1000 \g \cm^{-3}$. The mass and radius of the helium core are $M_{\rm core}=5M_{\rm \odot}$ and $R_{\rm core}=1.2R_{\rm \odot}$ respectively. At this stage the core is not dense enough for the NS to accrete at a high rate, and hence the onset of a CEE at this phase will not lead to r-process nucleosynthesis in our scenario. We therefore consider the case where the giant star swallows the NS during its second expansion. In this case the giant, if uninterrupted, reaches a maximum radius of about $600 R_{\rm \odot}$.  

The red giant might swallow the NS at any stage of this larger expansion, and so we examine the onset of the CEE at three different evolutionary points, as indicated in Table \ref{table:15Mproperties}. We assume that during the CEE the NS removes most of, or even the entire, envelope in a relatively short time. We therefore let the giant star evolve until it reaches the desired radius and we then remove the entire envelope at a constant rate during about $1000$ yr.

We note that if the removal time is too short and the envelope moves at a low velocity, the late jets that carry the r-process elements might collide with a heavy dense close wind and pass through a strong shock wave. The hot shocked r-process elements might disintegrate. We believe this is not likely because early jets will accelerate the envelope mass in the polar direction, such that the late jets that carry the r-process elements will not experience a strong shock. The opening of polar cones \citep{Sokeretal2019} implies that even shorter mass removal will not affect much our conclusions.

\begin{table}[H]
\begin{center}
 \begin{tabular}{|c c c c c c c |}
 \hline
$R$ & $M_{\rm core}$ & $R_{\rm core}$ & $E_{\rm bind}$ & $a_f(0.1)$ &  $a_f(0.3)$ & $a_f(1)$   \\ 
$(R_{\rm \odot})$ & $(M_\odot)$ & $(R_\odot$) & ($10^{50} \erg$) & $(R_\odot)$ &  $(R_\odot$) & $(R_\odot$) \\ 
 \hline
 150 & 2.5 & 0.09 & 1.06 & 0.006 & 0.02 & 0.06 \\
 300 & 2.5 & 0.09 & 0.97 & 0.007 & 0.02 & 0.07 \\
 600 & 2.6 & 0.08 & 0.59 & 0.01 & 0.04 & 0.12 \\
 \hline
\end{tabular}
\centering
\caption{Properties of the giant star during the CEE for the $M_{\rm ZAMS}=15M_{\rm \odot}$ stellar model. $R$ is the radius of the giant star when the NS penetrates its envelope, and $M_{\rm core}$ and $R_{\rm core}$ are the mass and radius of the CO core after the envelope is ejected, respectively. In all cases the envelope mass at the onset of the CEE is $M_{\rm env} = 8M_{\rm \odot}$. $E_{\rm bind}$ is the binding energy of the envelope (see  equation (\ref{eq:BindingEnergy}) in section \ref{sec:FinalOrb}). The last three columns list the final orbital separation after the envelope is removed, $a_f(\alpha)$, where $\alpha$ is the CEE parameter, between the core and the NS according to our assumptions that are described in section \ref{sec:FinalOrb}. }
\label{table:15Mproperties}
\end{center}
\end{table}

Following the same scheme for our $M_{\rm ZAMS}=30M_{\rm \odot}$, we find that this star reaches a radius of $R=53R_{\rm \odot}$ and a central density of $\rho \approx 350 \g \cm^{-3}$ at the peak of the first expansion phase. The mass and radius of the helium core are $M_{\rm core}=14M_{\rm \odot}$ and $R_{\rm core}=1.8R_{\rm \odot}$, respectively. 
For this stellar model the maximum radius in the second expansion phase is $R=1000R_{\rm \odot}$. In Table \ref{table:30Mproperties} we list the properties of the star at the three evolutionary points where we set a CEE. 
\begin{table}[H]
\begin{center}
 \begin{tabular}{|c c c c c c|}
 \hline
$R$ & $R_{\rm core}$ & $E_{\rm bind}$ & $a_f(0.1)$ &  $a_f(0.3)$ & $a_f(1)$ \\ 
$(R_{\rm \odot})$ & $(R_\odot$) & ($10^{50} \erg$) & $(R_\odot)$ &  $(R_\odot$) & $(R_\odot$) \\ 
 \hline
 200  & 0.18 & 3.16 & 0.008 & 0.02 & 0.07 \\
 600  & 0.17 & 2.91 & 0.008 & 0.02 & 0.08 \\
 1000 & 0.16 & 2.57 & 0.009 & 0.03 & 0.09 \\
 \hline
\end{tabular}
\centering
\caption{Like table \ref{table:15Mproperties} but for our $M_{\rm ZAMS}=30M_{\rm \odot}$ stellar model. In this case the mass of the envelope at the onset of the CEE is $M_{\rm env}=19M_{\rm \odot}$ and the mass of the core after the removal of the common envelope is  $M_{\rm core}=8.9M_{\rm \odot}$  }.
\label{table:30Mproperties}
\end{center}
\end{table}

As we can see in Figs. \ref{fig:MassDensity15M} and  \ref{fig:MassDensity30M}, when the NS penetrate the envelope at later stages, i.e. when the envelope of the giant star is larger, nuclear burning proceeds for a longer time before the entrance of the NS to the envelope, increasing a little the mass of the core. As a result, the core becomes denser and contracts. The binding energy of the envelope (see equation (\ref{eq:BindingEnergy})) decreases as well, facilitating the removal of the envelope mass by the NS.

\section{ORBITAL SEPARATION AT THE END OF THE CEE PHASE}
\label{sec:FinalOrb}

Let us estimate the final orbital separation between the NS and the CO core at the end of the CEE phase. The binding energy of the envelope before the NS is swallowed by the giant primary star is
\begin{equation}
E_{\rm bind}
=\frac{1}{2}\int_{M_{\rm core}}^{M_{\rm star}}\frac{Gm(r)}{r}dm,
\label{eq:BindingEnergy}
\end{equation}
where $M_{\rm star}$ and $M_{\rm core}$ are the total mass of the star and the mass of the CO core, respectively, and $m(r)$ is the mass coordinate of a stellar shell with radius r. The factor of half comes from the virial theorem, as we include the internal energy of the envelope. The values of the binding energies for our relevant models are listed in Tables \ref{table:15Mproperties} and \ref{table:30Mproperties}.  

The final orbital energy of the binary system after the ejection of the envelope is
\begin{equation}
E_{\rm orb,f}
=-\frac{\alpha GM_{\rm core}M_{\rm NS}}{2a_{\rm f}},
\label{eq:OrbitalEnergy}
\end{equation}
where $M_{\rm NS}=1.4 M_{\odot}$ is the mass of the NS in our scenario, $a_{\rm f}$ is the orbital separation of the system after the envelope is removed, and $\alpha$ is the efficiency parameter of the CEE (e.g., \citealt{LivioSoker1988}). 

Under the assumption of a canonical CEE (e.g., \citealt{Ivanovaetal2013}) the final orbital separation is determined by an equality between the released orbital energy and the envelope binding energy. Since in our case the magnitude of the final orbital energy of the system is much larger than the initial one, the orbital energy that the NS-core system releases is approximately given by equation (\ref{eq:OrbitalEnergy}), so that we take $E_{\rm orb,f} =-E_{\rm bind}$ and find 
\begin{equation}
a_{\rm f}
\simeq \frac{\alpha GM_{\rm core}M_{\rm NS}}{2E_{\rm bind}}.
\label{eq:afinal}
\end{equation}
We use equation (\ref{eq:afinal}) to calculate the orbital separation between the CO core and the NS at the end of the CEE for both of our stellar models and for three values of the CEE parameter $\alpha=0.1, 0.3$ and $1$. Equation (\ref{eq:afinal}) holds as long as the final orbital separation is larger than the radius of the core. If not, then after the NS enters the core we need to consider the removal of core material as well. This implies that the NS will continue to spiral-in even further. Namely, in cases where the NS enters to the core the final orbital separation given by equation (\ref{eq:afinal}) is an upper limit due to the onset of a second CEE between the NS and the core. As for our study we are only interested in examining whether the NS enters the core, these upper limits serve our goals.

\begin{figure}
\begin{center}
\vspace*{-0.9cm}
\includegraphics[width=0.49\textwidth]{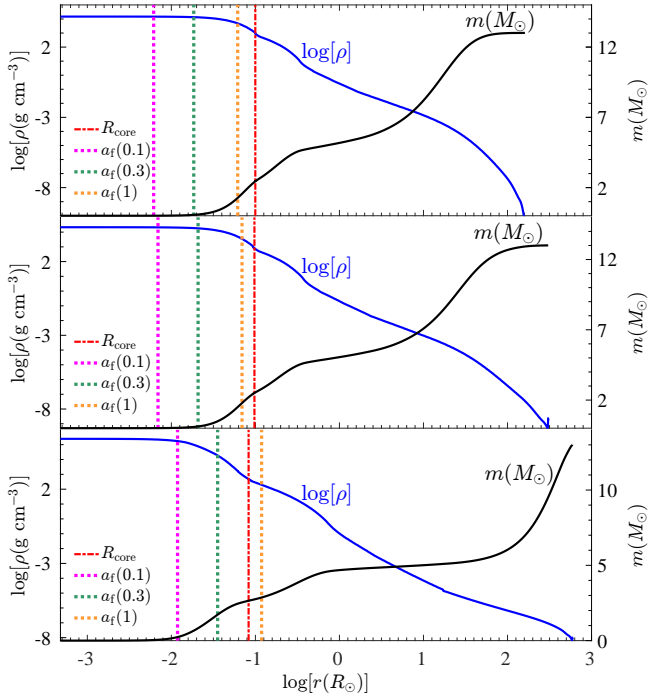}
\vspace*{-1.2cm}
\caption{The mass (black) and density (blue) profiles of our $M_{\rm ZAMS}= 15M_{\rm \odot}$ stellar model at the onset of the CEE assuming that the NS is swallowed by the giant star when it reaches a radius of $R=150R_{\rm \odot}$ (upper panel),$R=300R_{\rm \odot}$ (middle panel) or $R=600R_{\rm \odot}$ (lower panel). The red dashed-dotted line denotes the radius of the remaining CO core. The magenta (leftmost vertical line), green and orange dotted lines mark the orbital separation between the NS and the center of the core ($a_{\rm f}$) for $\alpha=0.1$, $\alpha=0.3$ and $\alpha=1$, respectively. 
}
\label{fig:MassDensity15M}
\end{center}
\end{figure}

\begin{figure}
\begin{center}
\vspace*{-0.85cm}
\includegraphics[width=0.49\textwidth]{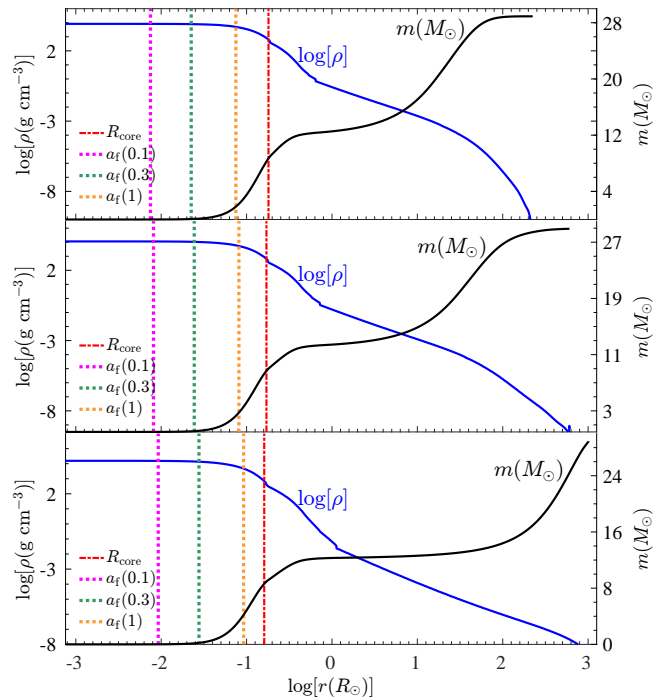}
\vspace*{-1.5cm}
\caption{Like Fig. \ref{fig:MassDensity15M} but for our $M_{\rm ZAMS}= 30M_{\rm \odot}$ stellar model and for cases where the giant swallows the NS when it reaches a radius of $R=200R_{\rm \odot}$ (Upper panel), $R=600R_{\rm \odot}$ (middle panel) or  $R=1000R_{\rm \odot}$ (lower panel).    
}
\label{fig:MassDensity30M}
\end{center}
\end{figure}

The values of the final orbital separation are listed in the last three columns of tables \ref{table:15Mproperties} and \ref{table:30Mproperties}, and are marked by vertical dashed lines in Figs. \ref{fig:MassDensity15M} and \ref{fig:MassDensity30M}. 

From Figs. \ref{fig:MassDensity15M} and \ref{fig:MassDensity30M} we can see that under our assumptions the NS enters the core in most of the cases. Only for the $15 M_\odot$ model that swallow the NS near the peak of its expansion and for high values of $\alpha$ the NS ends outside the core, yet very close to the core surface. If we would consider the jets that the NS is likely to launch while accreting some mass from the envelope (see section \ref{sec:intro} and \ref{sec:scenario}), then the final orbital separations would be somewhat larger. Numerical simulations show that jets are very efficient in removing envelope gas (e.g., \citealt{Shiberetal2019}). \cite{Sokeretal2019} mimic this effect by taking $\alpha>1$, e.g., $\alpha=3$. This means that in a large fraction of the cases that we find here, in which the NS enters the core, it might actually survive.

We conclude that for stars with initial mass in the general mass range of $\approx 10 M_\odot - 20 M_\odot$ when the NS is swallowed at the beginning of the second expansion phase it enters the core, while if it is swallowed later on it is likely to survive. In cases where the NS survives, after the explosion of the core a bound NS binary system might be formed (e.g., \citealt{VignaGomezeal2018}), that latter can merge (Fig. \ref{fig:CEJSN}). For more massive stars, $\ga 25 M_\odot$, the NS is likely to enter the core in more cases. In the present study we focus on cases where the NS enters the core.
 
\section{Inside THE CORE}
\label{sec:AcrretionRate}
\subsection{Accretion rate}
\label{subsec:AcrretionRate2}
When the NS enters to the core it starts accreting mass at a very high rate due to the high density in the core, most likely through an accretion disk. To form an accretion disk the specific angular momentum of the accreted matter should be larger than the angular momentum of a Keplerian motion on the companion's equator \citep{Soker2004}. Because of the density gradient in the core the accreted gas has angular momentum, and since the radius of the NS,  $\simeq10^6 \cm$, is much smaller than both the density scale height and the orbital separation, $\approx 10^8 \cm - 10^{10} \cm$, we expect that the gas is accreted on to the NS through an accretion disk (equation (7) in \citealt{Soker2004}).

 We take the orbit of the NS to be circular as it enters the core. Even if the NS had an eccentric orbit outside the envelope, due to strong tidal interaction and gravitational drag during the CEE, the orbit will become circular as we find in post-CEE binaries (e.g. \citealt{Zahn1977}; \citealt{Ivanovaetal2013}). 

The Bondi-Hoyle-Lyttleton mass accretion rate (\citealt{HoyleLyttleton1939}; \citealt{BondiHoyle1944}) inside the core is 
\begin{equation}
\dot M_{\rm BHL} 
\simeq \pi R_{\rm BHL}^{2}\rho_{\rm c} v_{\rm rel}
\label{eq:MdotBHL}
\end{equation}
where $\rho_{\rm c}$ is the density of the core in the location of the NS, $v_{\rm rel}$ is the velocity of the NS relative to the core, 
\begin{equation}
 R_{\rm BHL} 
 = \frac{2GM_{\rm NS}}{v_{\rm rel}^2 + c_{\rm s}^2}
 \approx \frac{2GM_{\rm NS}}{v_{\rm kep}^2},
 \label{eq:RBHL}
 \end{equation}
is the accretion radius, and $c_{\rm s}$ is the sound speed inside the core at the location of the NS. In the second equality of equation (\ref{eq:RBHL}) we made two simplifying assumptions, as is done in many similar cases, that have opposite effects on the accretion radius. (1) We neglected the sound speed in the denominator of the expression. This assumption increases the accretion radius. 
 (2) We assumed $v_{\rm rel} \approx v_{\rm kep}$ where $v_{\rm kep}$ is the orbital velocity of the NS inside the core. Since the core rotates $v_{\rm rel}$ is smaller than the Keplerian velocity and this assumption decreases the calculated accretion radius as given in equation (\ref{eq:RBHL}). With these two assumptions we find the mass accretion rate onto a NS spiraling-in inside the core of a giant massive star to be
\begin{equation}
 \dot M_{\rm BHL} 
 \simeq \frac{4 \pi G^2 M_{\rm NS}^2 \rho_{\rm c}}{v_{\rm kep}^3} .
 \label{eq:MdotBHLapprox}
 \end{equation}
 
There are two additional effects that alter the accretion rate and have to be considered in this scenario. Firstly, the jets remove mass from the NS vicinity as they expand out. This reduces somewhat the accretion rate. However, most of this material is perpendicular to the equatorial plane, and most of the accretion takes place from near the equatorial plane, so the effect of the jets on mass accretion is not large. On the other hand, the dense gas near the NS losses energy by neutrino cooling. This process reduces the pressure near the NS, hence increasing the accretion rate relative to $\dot M_{\rm BHL}$. In CCSNe, for example, neutrino cooling facilitated high accretion rate onto the newly born NS until the explosion occurs. However, since in the CEJSN r-process scenario the NS is cold, neutrino cooling has a less prominent effect. Overall, we assume that the BHL accretion rate as given by equation (\ref{eq:MdotBHLapprox}) is adequate for the goals of the present research.

\begin{figure}
\begin{center}
\vspace*{-2.70cm}
\includegraphics[width=0.49\textwidth]{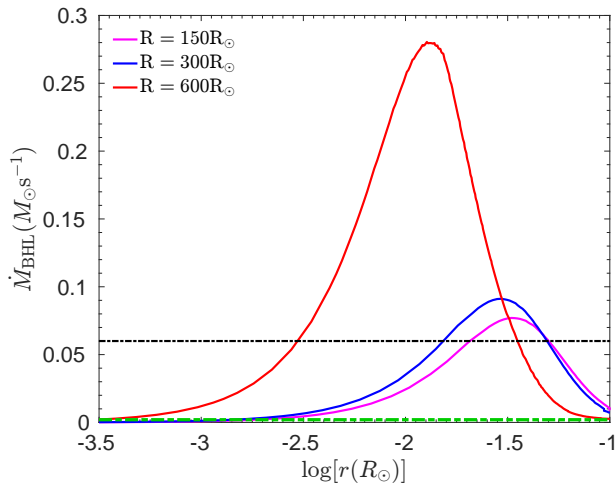}
\vspace*{-2.5cm}
\caption{Mass accretion rate onto the NS inside the CO core through an accretion disk for our $M_{\rm ZAMS}=15M_{\rm \odot}$ stellar model assuming the giant swallows the NS when the giant star reaches a radius of $R=150R_{\rm \odot}$ (magenta), $R=300R_{\rm \odot}$ (blue) or $R=600R_{\rm \odot}$ (red). The black dashed-dotted upper horizontal line denotes the accretion rate that gives a neutron to proton ratio of $N_{\rm n}/N_{\rm p} \simeq 5$ according to the results of \cite{kohri2005}. The lower horizontal green thick dashed-dotted line gives the threshold accretion rate for r-process nucleosynthesis according to \cite{Siegeletal2019}. 
}
\label{fig:15MAccretion}
\end{center}
\end{figure}

\begin{figure}
\begin{center}
\vspace*{-2.70cm}
\includegraphics[width=0.49\textwidth]{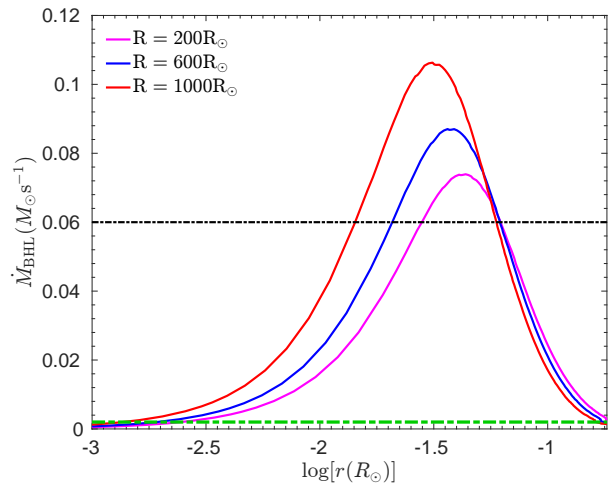}
\vspace*{-2.5cm}
\caption{Like Fig.\ref{fig:15MAccretion} but for the $M_{\rm ZAMS}=30M_{\rm \odot}$ model assuming the radius of the giant is $R=200R_{\rm \odot}$ (magenta), $R=600R_{\rm \odot}$ (blue) or $R=1000R_{\rm \odot}$ (red) when the NS enters the envelope. }
\label{fig:30MAccretion}
\end{center}
\end{figure}

We present the mass accretion rate of a NS of $1.4 M_\odot$ that spirals-in inside the cores of our $M_{\rm ZAMS}=15M_{\rm \odot}$ and $M_{\rm ZAMS}=30M_{\rm \odot}$ stellar models in figures \ref{fig:15MAccretion} and \ref{fig:30MAccretion}, respectively. 

When the mass of the core inner to the orbital location of the NS is about equal to the mass of the NS, We expect that the NS will gravitationally destroy the core, and the core material will form a thick disk around the NS (Fig. \ref{fig:CEJSN}). Although in this phase equation (\ref{eq:MdotBHLapprox}) does not apply anymore, we note that the accretion rate from the thick disk might be larger even.  However, heavy hydrodynamical simulation are required to find the actual accretion rate from the disk.

To analyse the accretion rate in the scope of our paper, lets us assume that the upper limit of the accretion rate is obtained at the radius in which the core is destroyed and equation (\ref{eq:MdotBHLapprox}) ceases to apply. From Figs. \ref{fig:MassDensity15M} and \ref{fig:MassDensity30M} we can find the radius for which the core is destructed, and infer the maximum accretion rate by using Figs. \ref{fig:15MAccretion} and \ref{fig:30MAccretion}, respectively. 

\cite{Winteleretal2012} study the nucleosynthesis of heavy r-process elements inside jets launched from a NS in MHD-driven supernovae. They find it is possible to reproduce the second and third peaks of the r-process (i.e. heavy r-process) when the parameter $Y_e \equiv N_e/(N_p+N_n) \la 0.15$, where $N_e$, $N_p$ and $N_n$ are the electron, proton and neutron density, respectively.
This corresponds to a neutron to proton ratio of $N_n/Np \ga 5$. 
Using figure 6b from \cite{kohri2005} we find that the minimum value of accretion rate needed to obtain this neutron-rich outflow is about $\dot M_{\rm nr} \simeq 0.06 M_\odot \s^{-1}$, as we mark by an horizontal black dashed-dotted line in Figs. \ref{fig:15MAccretion}- \ref{fig:30MAccretion}.
As can be seen in the figures, in half of the cases the upper boundary of the accretion rate in the CEJSN r-process scenario is high enough for heavy r-process nucleosynthesis. In the rest of the cases, the accretion rate is about half of the accretion rate needed to produce heavy r-process elements. However, as we expect the accretion rate to be higher from the BHL limit due to the formation of a thick accretion disk from the destructed core material, we conclude that in even more cases the neutron to proton ratio is high enough for the formation of the heavies elements.
 
In their recent study \cite{Siegeletal2019} find the threshold accretion rate to form a neutron-rich outflow to be much lower, at about $\dot M_{\rm nr} \simeq 0.002 M_\odot \s^{-1}$. We mark this threshold by a thick dash-dotted green line in Figs. \ref{fig:15MAccretion}- \ref{fig:30MAccretion}.   We note that for all our scenarios the accretion rate is above 10 times larger than the accretion rate required to form heavy r-process elements according to the calculations of \cite{Siegeletal2019}, even without taking into account the formation of the thicker accretion disk that accretes material faster.

We run our stellar models for lower ($Z = 0$; stars of the first generation) and higher ($Z=0.005$) metalicities to examine how the progenitors metalicity affects our results. We found that for giant stars of the first stellar generation, the results do not differ by much from the cases we studied above. For the models in which the metalicity is higher, however, 
the NS enters the core in most of the cases, yet the accretion rate is too low for r-process nucleosynthesis in the vast majority of the NS-core merger events. This means that even though that the binary frequency among the population of massive stars does not depend on metalicty (i.e. \citealt{Eldridgeetal2017} and references therein), it seems that the rate of CEJSN r-process scenario decreases with time, stressing the importance of these events in the early universe).

We safely conclude that the accretion rate in the CEJSN r-process scenario is high enough to form the neutron-rich outflow that is required for the heavy r-process nucleosynthesis.   

\subsection{Initial conditions}
\label{subsec:Conditions}
\subsubsection{Qualitative justifications}
\label{subsubsec:QualitieveJustification}
We use here the results of \cite{kohri2005} and \cite{Siegeletal2019} to argue for the formation of a neutron-rich (low electron fraction) outflow in the jets. We assume that by using the accretion rate we calculate, we can rely on these papers
to deduce the neutron to proton ratio in our scenario. Let us justify this assumption by looking at 4 properties of the system. 

 (1) \textit{Accretion disk.} Both papers take the accretion onto the central object to be from an accretion disk. In our case the accretion must also occur via an accretion disk, as explained in section \ref{sec:scenario} and in the begining of subsection \ref{subsec:AcrretionRate2}. 

 (2) \textit{Nuclear statistical equilibrium.} In the papers mentioned above, a newly-formed, very hot proto-neutron star is considered, surrounded by shocked material. This material flows to the accretion disk where it reaches nuclear statistical equilibrium \citep{kohri2005, Siegeletal2019}. Although in our case the accreted mass starts as CO rich material, when it flows inside the accretion disk, due to the high number density and high energy density, it also reaches nuclear statistical equilibrium, i.e., the material `foregts' its initial composition.  

 (3)  \textit{Formation of a neutron rich gas.} The very high density in the accretion flow leads to high electron degeneracy, such that the electrons have sufficient energy to convert protons to neutrons by the process $e^{-} + p \rightarrow n + \nu_e$, and a neutron-rich gas is obtained.  Since our flow is similar in mass accretion rate and flow geometry to that of  \cite{kohri2005} and \cite{Siegeletal2019}, we expect the accretion disk in the CEJSNe scenario, hence the jets that it launches, to be neutron-rich.

(4) \textit{Neutrino flux.}  {{{{{{ In the case of a CCSN the newly born NS is very hot and emits a huge amount of energy in neutrinos. After several seconds the emission substantially decreases. In our scenario the NS starts cold and does not emit neutrinos. }}}}}} The accretion disk, on the other hand, cools down by neutrino emission due to the relatively high temperature of the disk. Since it does not accrete much before it launches jets, the neutrino flux from cooling is much below that in the calculations of \cite{kohri2005} and \cite{Winteleretal2012}. Considering that neutrinos convert back neutrons to protons via the reaction $\nu_e + n \rightarrow p + e^{-}$, our conditions {{{{{{might actually be}}}}}} more favourable for r-process than in these previous studies.
{{{{{{ In the work \cite{Siegeletal2019} the accretion is around a black hole, so neutrino cooling comes only from the accretion disk, and plays a major role in the accretion process. Our case includes some uncertain parameters, but nonetheless seems to be somewhat similar as we show below.  }}}}}}

\subsubsection{Quantitative justifications}
\label{subsubsec:QantitativeJustification}
 
 
{{{{{{ We quantitatively justify points (2) and (4) from section \ref{subsubsec:QualitieveJustification}. We first note that \cite{Siegeletal2019} use the standard $\alpha$-disk model \citep{ShakuraSunyaev1973} with $\alpha = 0.01$ to infer the typical density and temperature of the accretion disk in their collapsar scenario. As usual in the accretion disk model $\alpha= \nu /C_s H$, where $\nu$ is the coefficient of the kinematic viscosity, $C_s$ the sound speed and $H$ is the vertical scale height of the disk. We found that if we also use the standard accretion disk model with $\alpha=0.01$ and with the typical properties of our accretion disk, we derive density and temperature of the same order of magnitude as in \cite{Siegeletal2019}.
Nevertheless, in the following calculations we go one step further and use expressions that already include neutrino cooling.  
 }}}}}}

{{{{{{ We start with the calculations by \cite{Chevalier1996} of a neutrino-cooling dominated accretion disk. We scale his equations to the properties of our system to find the typical temperature and density in the disk. We take the radius inside the accretion disk to be $r = 50 \km$ as we expect it to extend from the NS surface at $r \simeq 12 \km$ to several times this radius. We scale the mass accretion rate with $\dot M = 0.05M_{\rm \odot}\s ^{-1}$ according to Figs. \ref{fig:15MAccretion} and \ref{fig:30MAccretion}. \cite{Chevalier1996} limits the value of $\alpha$ for a neutrino-cooling dominated disk. Scaling with our parameters in his equation gives }}}}}}
\begin{equation}
\begin{split}
\alpha  \lesssim  0.04   
&
\left( \frac{\dot M}{0.05 M_\odot \s^{-1}} \right)^{0.53}  
\left(\frac{M}{1.4 M_\odot} \right)^{0.29}
\\
& 
\times 
\left(\frac{r}{50 \km} \right)^{-0.87}
\end{split}
\label{eq:alpha}
\end{equation}
Below we take $\alpha = 0.01$. {{{{{{{ The uncertainty in the value of $\alpha$ introduces uncertainty in the quantities we derive below. Less significant is the uncertainty in the value of the radius in the disk. As stated, we scale the radius inside the disk with about four times the radius of the NS. The radius can have smaller values, down to about $12 \km$ (the NS itself), or be somewhat larger than $r=50 \km$. Even for a radius of $150 \km$ our assumptions still hold, as the radius is raised to a power of $<1.5$ in the different expressions (when the dependence of $\alpha$ on the radius is included). Basically, the relevant outflow from the disk will take place in the range $r \simeq 12 -100 \km$, and so we use $r=50 \km$ as scaling value. }}}}}}}

Scaling now the density and temperature from \cite{Chevalier1996} we obtain \begin{equation}
\begin{split}
\rho \approx 6\times 10^{10} &  
\left( \frac{\dot M}{0.05 M_\odot \s^{-1}} \right)^{0.84}  
\left(\frac{M}{1.4 M_\odot} \right)^{0.76}
\\
& 
\times 
\left(\frac{r}{50 \km} \right)^{-2.29}
\left(\frac{\alpha}{0.01} \right)^{-1} \g \cm^{-3} , 
\end{split}
\label{eq:density}
\end{equation}
and
\begin{equation}
\begin{split}
T \approx  5.5\times 10^{10}   &
\left( \frac{\dot M}{0.05 M_\odot \s^{-1}} \right)^{0.11}  
\left(\frac{M}{1.4 M_\odot} \right)^{0.16}
\\
& 
\times 
\left(\frac{r}{50 \km} \right)^{-0.47} \K .
\end{split}
\label{eq:temperature}
\end{equation}
{{{{{{ For the vertical scale height of the accretion disk we also scale an expression from \cite{Chevalier1996} and find that }}}}}}
\begin{equation}
\begin{split}
\frac{H}{r} \approx 0.42   
\left( \frac{\dot M}{0.05 M_\odot \s^{-1}} \right)^{0.05} & 
\left(\frac{M}{1.4 M_\odot} \right)^{-0.42}
\\
& 
\times 
\left(\frac{r}{50 \km} \right)^{0.26}.  
\end{split}
\label{eq:Thickness}
\end{equation}
{{{{{{ This implies that the disk is not very thin.   }}}}}}

{{{{{{ From the vales in equations (\ref{eq:alpha})-(\ref{eq:temperature}) we can calculate the typical  neutrino-cooling timescale in the disk. Taking the neutrino cooling rate as \citep{BrownWeingartner1994} $\epsilon_{\rm \nu}  =2.3\times 10^{31} \left(\frac{T}{5.5 \times 10^{10} \K} \right)^{9} \erg \cm ^{-3} \s ^{-1}$, we calculate the neutrino cooling timescale }}}}}}
\begin{equation}
t_{\rm \nu,cool} 
 \simeq \frac{aT^{4}}{\epsilon_{\rm \nu}} \approx 3\times 10^{-3} \left(\frac{T}{5.5 \times 10^{10} \K}\right)^{-5} \s
\label{eq:NeutrinoCoolingTime}
\end{equation} 
{{{{{{ We calculate the radial velocity in the disk from mass accretion rate 
$\dot M = 2 \pi r 2 H \rho v_r$, and find it to be is $v_r \approx 100 \km \s^{-1}$. The inflow time of the gas in the disk from $r \simeq 50 \km$ is $t_{\rm in} \approx r/v_r \approx 0.5 \s \gg t_{\rm \nu,cool}$. Namely, neutrino cooling in the disk is important. }}}}}} 

{{{{{{ From the above results we can safely conclude that our derivations are self-consistent, and that neutrino cooling plays a major role in the accretion disk of our scenario. Neutrino cooling lowers the temperature of the accreted gas, reducing the pressure, and ensuring a mildly-electron degenerate state, as in the scenario of \cite{Siegeletal2019}. }}}}}} 

{{{{{{ From \cite{Beloborodov2003} the electrons become degenerate below the characteristic degeneracy temperature of }}}}}} 
\begin{equation}
T_{\rm deg} = 7.5 \times 10^{10} \left( \frac{\rho}{6 \times 10^{10} \g \cm^{-3}} \right)^{1/3} \K .
\label{eq:DegeneracyTemperature}
\end{equation}
{{{{{{ This is somewhat higher than the typical temperature we find in equation (\ref{eq:temperature}), implying that we have a mildly degenerate electron gas as \cite{Siegeletal2019} find in their scenario.}}}}}}
{{{{{{Moreover, according to Figure. 1 of  \cite{Beloborodov2003}, the baryonic matter in the accretion disk of our scenario is in nuclear statistical equilibrium and is dominated by free nucleons with electron to baryon ratio of $Y_e < 0.3$, i.e., a large fraction of neutron to protons.. }}}}}}

{{{{{{ Overall, we conclude that the accretion disk conditions in the CEJSN r-process scenario resemble the conditions from the studies of \cite{kohri2005} and \cite{Siegeletal2019}. We expect then the outflow properties to be quite similar.  }}}}}}


\subsection{Nucleosynthesis in jets}
\label{subsec:Nucleusynthesis}

As we mention above, at the stage of high mass accretion rate the core mass inner to the orbit of the NS is $\approx 1 M_\odot$. A large fraction of this mass is expected to be accreted at this high rate. As described in section \ref{sec:scenario}, \cite{Papishetal2015} estimated that the jets in the CEJSN r-process scenario are expected to synthesis a mass of $\approx 0.001-0.01 M_\odot$ of heavy r-process elements per event.
  This is based on that a fraction of $\approx 10 \%$ of the accreted mass is launched in the jets and $\approx 10\%$ of the mass in the jets is transferred to heavy r-process elements.  
 If we would have used the recent results of \cite{Siegeletal2019} the mass of the r-process elements would have been higher even. \cite{Siegeletal2019} find the ejected mass to be $\approx 30 \%$  of the accreted mass. Under this assumption we can take a more optimistic yield than what \cite{Papishetal2015} used, namely, we can take the yield per CEJSN r-process event to be $\approx 0.01-0.03 M_\odot$ of heavy r-process elements. 

As we mentioned in section \ref{sec:scenario}, \cite{Chevalier2012}  
estimated the rate of events in which a NS enters the envelope of a giant star to be $\approx 1 \%$ of the total rate of CCSNe. \cite{Papishetal2015} found that if this is also the rate of CEJSN r-process events, then combining it with the yield of r-process elements that they derived we obtain the observed amount of heavy r-process elements. Since we take it more likely that the yield of r-process elements per event be $\approx 10$ times higher than the one mentioned in \cite{Papishetal2015}, we obtain that the number of events that synthesis r-process elements is $\approx$ 10 times lower, i.e., $\la 10^{-3}$ of the total number of CCSNe. {{{{}}}}

We can reach a similar conclusion from another direction. 
\cite{Siegeletal2019} estimate that a total mass of  $\approx 0.2-1 M_\odot$ 
should be accreted per collapsar (or per long gamma ray burst) to account for the solar system r-process abundances. Since the basic accretion process and accreted mass of a collapsar in the scenario of \cite{Siegeletal2019} is as the one of the CEJSN r-process studied by \cite{Papishetal2015} and here, we expect a similar r-process yield per event.
Considering the rate of CCSNe is about $\approx 10^3$ times that of long gamma ray bursts \citep{WandermanPiran2010}, we conclude that the number of CEJSN r-process events should be $\approx 10^{-3}$ that of CCSNe to explain the solar system r-process abundances by the CEJSN r-process scenario.  This is compatible with the rate that \cite{Beniaminietal2016b} deduce from their study of Eu abundance in dwarf galaxies.   Based on the estimate of \cite{Chevalier2012} then, about one in ten events of a NS that enters the envelope of a giant star should end as CEJSN r-process event. 

We conclude that the CEJSN r-process rate is $\approx 10\%$ of the NS-NS merger rate. Therefore, most cases in which a NS enters A CEE phase end with the formation of two NSs, and only the minority ends with a NS that enters the core. This ratio is smaller than the number of core-NS merger cases we find here for two reasons (see discussion in section \ref{sec:FinalOrb}). (1) We simulated mainly cases that might end in core-NS merger. (2) We neglected the removal of envelope mass by jets, e.g., \cite{Sokeretal2019}, that will prevent merger even in cases that here we find that end in merger.

\section{DISCUSSION AND SUMMARY}
\label{sec:summary}

We studied the CEJSN r-process scenario following \cite{Papishetal2015}. In this scenario a NS  spirals-in through the envelope of a massive giant star and into the core, accretes mass from the core and launches jets. In the last phase of the CEE the NS destroys the core and the core material turns to a thick accretion disk from which the NS accretes mass at a high rate \citep{Papishetal2015}. Due to the large amount of accreted mass, the NS might turn into a black hole. The r-process nucleosynthesis takes place inside the jets in this last phase. 
We describe the evolutionary route of the CEJSN r-process scenario in section \ref{sec:scenario} and schematically in the left branch of Fig. \ref{fig:CEJSN}. 
 
We followed the evolution of two low-metalicity massive stars until the giant phase, and assumed that the giant swallows a NS companion and the system enters a CEE phase (section \ref{sec:NumericalScheme}).  
In section \ref{sec:FinalOrb} we used the canonical common envelope prescription to estimate the final orbital separation of the NS from the center of the core. 
We found that in several cases the NS falls into the core of the giant star before its envelope is fully ejected, as can be seen in Figs. \ref{fig:MassDensity15M}-\ref{fig:MassDensity30M}. We also found that in cases where the giant star swallows the NS the first time it expands the core is not dense enough for the NS to accrete at a high enough rate that can lead to r-process nucleosynthesis. The CEJSN r-process scenario requires the NS to enter the envelope of the giant star during the second expansion phase of the giant. 
      
We then used the Bondy-Hoyle-Lyttleton accretion rate to estimate the mass accretion rate onto a NS spiraling-in inside the core of a giant massive star under our assumptions (section \ref{sec:AcrretionRate}). Due to the small radius of the NS the accretion is through an accretion disk, which we assume launches jets (or bipolar disk winds). As seen in Figs. \ref{fig:15MAccretion} and \ref{fig:30MAccretion}, in many cases the mass accretion rate is higher than the minimum value needed for the formation of highly neutron-rich material that leads to r-process nucleosynthesis.  On these figures we drew two different thresholds for this mass accretion rate. The upper one is from the results of \cite{kohri2005} where the line indicates the accretion rate for which we find that the neutron to proton ratio is about $N_n/N_p \simeq 5$, and the lower one is from the recent results of \cite{Siegeletal2019}. For both thresholds the CEJSN r-process scenario provides a sufficiently high accretion rate onto the NS to account for the formation of the second and the third r-process peaks (heavy r-process). 

The amount of mass that is accreted onto the central object, a NS or a black hole, in the new study of \cite{Siegeletal2019} is similar to that onto the NS in the CEJSN r-process scenario \citep{Papishetal2015}. However, the r-process yield that \cite{Siegeletal2019} derive is about an order of magnitude higher than the one \cite{Papishetal2015} estimated. Using the same assumptions as in \cite{Siegeletal2019} the new value of r-process elements per CEJSN r-process event amounts to $\approx 0.01-0.03 M_{\rm \odot}$ 
 This requires that one CEJSN r-process event occurs per about one thousand CCSN events if the CEJSN r-process scenario is the major contributer to r-process nucleosynthesis. Based on the estimation by \cite{Chevalier2012} we conclude that $\approx 10 \%$ of the systems where a NS enters the envelope of a giant star should end as CEJSN r-process events. 
 
We list two types of observations that might strengthen the general CEJSN scenario, examine the CEJSN frequency and compare it with our estimations. Indirectly, some peculiar and rare supernovae of massive stars might indicate to the CEJSN mechanism \citep{Sokeretal2019}. However, for each of the peculiar supernovae other models exist, so there is no strong hint from those cases yet. The detection of unique gravitational waves signatures from the merger of a NS with the compact core of a giant star can serve as an evidence to a CEJSN event. \cite{Nazinetal1997} studied a similar process for gravitational waves emission. In a recent study,  \cite{Ginatetal2019} ) present detailed calculations and argue that next-generation space-based gravitational wave detectors will be able to detect gravitational waves from NS-core merger. 
 
There are several open questions that only 3D hydrodynamical simulations can answer. One of them is the question whether neutrino from the cooling accreted mass will not convert too many neutrons back to protons. The new results of \cite{HaleviMosta2018} and \cite{Mostaetal2018} make us optimistic that in most cases the neutrino flux will not prevent r-process nucleosynthesis, although it will influence the outcome. \cite{HaleviMosta2018} find in some cases robust r-process nucleosynthesis, as long as electron and anti-electron neutrino luminosities are $L_{\nu} + L_{\bar \nu} < 10^{53} \erg \s^{-1}$. For our average accretion rate of $0.06 M_\odot \s^{-1}$ and less, the average neutrino luminosity, including all types, is $L_{\nu, {\rm tot}}<2 \times 10^{52} \erg \s^{-1}$. Namely, $L_{\nu} + L_{\bar \nu}$ will be lower even. 

We went one step beyond the preliminary suggestion of the CEJSN r-process scenario \citep{Papishetal2015}.
At present this scenario has some assumptions and one speculative component, yet also some calculations that stress its merit. In Table \ref{table:Summary} we list our assumptions along supporting arguments that show the assumptions and the speculation stand on solid grounds, despite the open questions that are left for further investigation. In the last column we list the future calculations required for a further development of this scenario.
\begin{table*}
\caption{The status of the different phases of the CEJSN r-process scenario}
\centering
\begin{tabular}{|  c | c | c |  }
\hline
Process  & {\textcolor[rgb]{0.00,0.59,0.00}{Calculation/}} &  Supporting arguments        \\ 
         & {\textcolor[rgb]{0.8,0.0,0.8}{Assumption/}}  &          \\ 
         & {\textcolor[rgb]{0.98,0.00,0.00}{Speculation}}  &        \\ 
\hline
          \hline   
NS merges    & {\textcolor[rgb]{0.00,0.59,0.00}{We showed this}}  & Gravitational waves show NSs merger, implying                                                 \\ 
with the core& {\textcolor[rgb]{0.00,0.59,0.00}{occurs (Figs. \ref{fig:MassDensity15M}+\ref{fig:MassDensity30M})}}& many NS binaries end CEE with a very small  \\ 
             &                 & separation. In the tail of this distribution                                                                                     \\ 
             &                 & we expect NSs to merge during the CEE.                                                                                          \\ 
\hline   
Accretion     & {\textcolor[rgb]{0.8,0.0,0.8}{Accretion rate $\dot M_{\rm acc}$ }}& (1) High accretion rates in some CEE progenitors     \\ 
on to the NS  & {\textcolor[rgb]{0.8,0.0,0.8}{at the  BHL rate}}                  & of PNe (BL14, Sa17);  (2) Studies show jets allow     \\ 
              &                                                                   &  high accretion rates (Sh16; C18).                     \\ 
\hline   
Accretion    & {\textcolor[rgb]{0.8,0.0,0.8}{The accreted mass}} & The very small radius of the NS     \\ 
disk in CEE  & {\textcolor[rgb]{0.8,0.0,0.8}{forms a disk}}      & (scaled by eq. 7 in So04).     \\ 
          \hline   
Destroyed core& {\textcolor[rgb]{0.8,0.0,0.8}{We argued the flow}}  & Based on 3D simulations of a NS tidally      \\ 
forms a       & {\textcolor[rgb]{0.8,0.0,0.8}{is similar to that}}  & destroying a white dwarf, and showing the    \\ 
massive disk  & {\textcolor[rgb]{0.8,0.0,0.8}{in collapsars (Si19)}}& formation of an accretion disk (e.g., Bo17).      \\ 
\hline   
Accretion rate & {\textcolor[rgb]{0.00,0.59,0.00}{We showed the BHL}}   & The BHL accretion rate is $\approx 1.2-4$ and $>10$   \\ 
high enough for& {\textcolor[rgb]{0.00,0.59,0.00}{accretion rate meets}} & times the rate that the r-process requires for  \\ 
neutron-rich & {\textcolor[rgb]{0.00,0.59,0.00}{this limit (Figs. \ref{fig:15MAccretion}+\ref{fig:30MAccretion})}}& accretion on to a NS (Ko05) and a BH (Si19),  \\ 
disk           &                                                         &  allowing us a high margin of uncertainty.     \\ 
          \hline   
Jets     & {\textcolor[rgb]{0.8,0.0,0.8}{The disk launches}} & Energetic jets in some PNe (BL14, Sa17).     \\ 
launching& {\textcolor[rgb]{0.8,0.0,0.8}{jets in CEE}}  &   \\ 
                                                                                                           
          \hline   
R-process   & {\textcolor[rgb]{0.98,0.00,0.00}{We did not }}         & Based on our finding/assumption                                     \\ 
inside jets & {\textcolor[rgb]{0.98,0.00,0.00}{ calculate r-process}}& that the accretion flow is similar to that in                       \\ 
            & {\textcolor[rgb]{0.98,0.00,0.00}{}}                    & collpasers, where Si19 find heavy r-process.                          \\ 
\hline   
Neutrino flux    &{\textcolor[rgb]{0.00,0.59,0.00}{We calculated total}}& HM18 found r-process nucleosynthesis for                          \\ 
does not prevent &{\textcolor[rgb]{0.00,0.59,0.00}{neutrino flux $L_{\nu, {\rm tot}}$}}& $L_{\nu} + L_{\bar \nu} < 10^{53} \erg \s^{-1}$,    \\ 
r-process in jets&{\textcolor[rgb]{0.00,0.59,0.00}{$<2 \times 10^{52} \erg \s^{-1}$ }}   & a condition the CEJSN r-process obeys.        \\ 
 \hline   
Rate of & {\textcolor[rgb]{0.00,0.59,0.00}{We require }}       & From the estimate by Ch12 of rate of events      \\ 
CEJSNe  & {\textcolor[rgb]{0.00,0.59,0.00}{1 CEJSN r-process }}& where a NS enters an envelope of a giant,      \\ 
        &{\textcolor[rgb]{0.00,0.59,0.00}{per $\approx 1000$ CCSNe;}}& we found that it is sufficient that only       \\ 
        & {\textcolor[rgb]{0.8,0.0,0.8}{We assume this}}       & $\approx 10 \%$ of these events end as     \\ 
        & {\textcolor[rgb]{0.8,0.0,0.8}{rate is feasible}}      &  CEJSN r-process events.     \\ 
          \hline   
   \hline   
\end{tabular}
\footnotesize
\newline
\textbf{Acronyms:} BH: black hole; BHL: Bondi-Hoyle-Lyttleton; CEE: common envelope evolution;  NS: neutron star; PNe: planetary nebulae; YSOs: young stellar objects. 
\newline
\textbf{References:} Bo17: {Bobrick2017}; BL14: \cite{BlackmanLucchini2014}; C18: \cite{Chamandyetal2018}; Ch12: \cite{Chevalier2012}; HM18: \cite{HaleviMosta2018}; Ko05: \cite{kohri2005}; Sa17: \cite{Sahaietal2017}; Sh16: \cite{Shiberetal2016}; Si19: \cite{Siegeletal2019}; So04: \cite{Soker2004}.
\vspace*{0.5cm}
\label{table:Summary}
\end{table*}

We will not discuss in great length all the details in Table \ref{table:Summary}, as we present additional references the reader can turn to for further analysis. We here elaborate, as an example, on the two rows of the accretion rate and disk formation (more on these points are in earlier papers on the CEJSN, e.g., \citealt{Sokeretal2019} and \citealt{Gilkisetal2019}). 
This example emphasises the usage of results from other astrophysical objects, planetary nebulae in this case. 
 
Studies reach different conclusions on the accretion rate by the more compact object that spirals-in inside a giant envelope and on whether the accreted gas forms an accretion disk around the compact object  
(e.g., \citealt{RasioShapiro1991, Fryer1996, Lombardietal2006, RickerTaam2008, MacLeod2015, MacLeod2017}). 
Although there are claims for low accretion rates, e.g., \cite{MacLeodRamirezRuiz2015b}, there are several new studies that suggest that the accretion rate inside a common envelope can be close to the BHL value. The key process is that jets remove energy and angular momentum from the vicinity of the accreting object, and by that allow the high accretion rate (e.g., \citealt{Shiberetal2016, Chamandyetal2018}). \cite{BlackmanLucchini2014} and \cite{Sahaietal2017} argue that the large momenta in some bipolar planetary nebulae must come from jets that a main sequence companion launches inside a common envelope while accreting at a high rate. Because of the very small radius of a NS, it is much easier for the accretion flow to form an accretion disk around a NS than around a main sequence star. 
This discussion strengthens our belief that the accretion rate onto the NS is close to the BHL accretion rate and that an accretion disk indeed forms, despite the need for 3D hydrodynamical numerical simulations for a more careful study of these aspects.  
From the formation of an accretion disk, and observations of many astrophysical objects where accretion disks launch jets, we also consider our assumption that the NS and its accretion disk launch jets to be robust. 

 There is only one ingredient of the scenario that is on the boarder between a speculation and an assumption. This is the question of whether the heavy r-process takes place inside our jets. First we note that we expect the accretion rate to be high enough to form a neutron-rich matter in the inner part of the accretion disk, both for accretion on to a NS, and more so for accretion on to a black hole (Figs. \ref{fig:15MAccretion} and \ref{fig:30MAccretion}).
Here our optimistic assumption/speculation that the heavy r-process does take place inside the jets is based on the results of \cite{Siegeletal2019} even that they consider jets from a black hole. But even before the more optimistic results of \cite{Siegeletal2019}, \cite{Papishetal2015} argued that earlier results of neutron-rich jets launched by the newly born NS in CCSNe  (e.g., \citealt{Winteleretal2012}) lends support to the CEJSN r-process scenario. We find that even this assumption/speculation is justified.


 Overall, we strengthened the suggestion made by \cite{Papishetal2015} that CEJSNe constitute a promising heavy r-process site in the early Universe. 
 But we admit, as we list in the last column of Table \ref{table:Summary}, that a large number of additional calculations is needed before we can claim that the CEJSN r-process scenario stands on a solid ground. These simulations should include 3D hydrodynamical simulations of the CEE, nuclear reactions to establish the formation of a neutron-rich accretion disk, and then the nucleosynthesis of r-process elements in the jets.
Finally, we will perform a population synthesis study to find whether there are enough CEJSN r-process events.

\section*{Acknowledgments}
We thank Avishai Gilkis, Oded Papish, and Efrat Sabach for helpful comments.  We thank an anonymous referee for very helpful and detailed comments that improved the presentation of our results and the physics involved. We acknowledge support from the Israel Science Foundation. A.G. was supported by The Rothschild Scholars Program- Technion Program for Excellence.

\end{document}